\documentclass[twocolumn,showpacs,preprintnumbers,amssymb]{revtex4}
\usepackage{pdfpages}
\usepackage{graphics}
\usepackage{epsfig} 
\usepackage{hyperref}
\usepackage{graphics}
\usepackage{color}      
\usepackage{graphicx}
\usepackage{graphicx}
\usepackage{dcolumn}

\begin{document}

\title{ Thermodynamic Features of a Heat Engine Coupled with Exponentially Decreasing Temperature Across the Reaction Coordinate, as well as Perspectives on Nonequilibrium Thermodynamics}
\author{Mesfin Asfaw  Taye}
\affiliation {West Los Angeles College, Science Division \\9000  Overland Ave, Culver City, CA 90230, USA}

\email{tayem@wlac.edu}


\begin{abstract}

In this study, we advance the understanding of non-equilibrium systems by deriving thermodynamic relations for a heat engine operating under an exponentially decreasing temperature profile. Such thermal   configurations  closely mimic spatially localized heating such as laser-induced thermal gradients. Using exact analytical solutions, we show that this arrangement results in significantly higher velocity, entropy production, and extraction rates than piecewise thermal profiles, while exhibiting reduced irreversibility and complexity relative to linear or quadratic gradients. We further examine the thermodynamic behavior of the Brownian particles in the networks. Our study reveals that the  velocity and entropy production rates remain independent of network size; on the contrary,  extensive quantities such as total entropy depend on  the number of microstates. Additionally, we show  that a Brownian particle in a ratchet potential with spatially varying temperature achieves directed motion, even without external forces driven by solely thermal asymmetry. These findings highlight the critical role of temperature asymmetry in controlling the transport processes and optimizing the particle dynamics. This in turn will have  promising applications in microfluidic devices and nanoscale sensors. Finally, we explore the influence of the system parameters on the efficiency and performance of the heat engine.  The  exponential temperature profiles enable faster velocities while simultaneously exhibiting higher  efficiency compared with other thermal arrangements. Moreover, by addressing key questions on entropy production,  we provide insights into the transition between nonequilibrium and equilibrium systems and contribute tools for optimizing energy-efficient systems in both natural and engineered settings.

\end{abstract}
\pacs{Valid PACS appear here}
\maketitle


 \section{Introduction}

Nonequilibrium thermodynamics and statistical mechanics remain  challenging because of the complexity of systems that are far from equilibrium. Unlike equilibrium systems, which follow well-defined principles, such as the Boltzmann distribution, nonequilibrium systems lack a universal framework. Key challenges include understanding the microscopic entropy production, predicting steady-state distributions, and addressing the impact of rare fluctuations. Although advances such as the thermodynamic uncertainty relation offer insights, a complete understanding of nonequilibrium processes remains incomplete, highlighting the need for better theoretical and experimental approaches.

 However, recent studies on nonequilibrium thermodynamic systems have garnered significant attention owing to their profound implications for science and technology \cite{mu1,mu2,mu3,mu4,mu5,mu6,mu7,mu8,mu9,mu10,mu11,mu12,ta1,mu13,mu14,mu15,mu16,mu17,muu17,mu25,mu26,mu27}. Research efforts have explored both classical \cite{mu17,muu17} and quantum systems \cite{mu25,mu26,mu27} with a particular focus on entropy production. The dependence of entropy production was investigated using the master equation approach for discrete systems \cite{mu1,mu2,mu3,mu4,mu5} and the Fokker-Planck equation for continuous systems \cite{mu7,mu8,ta1,muuu17,muuu177}. Various thermodynamic relations have been derived using stochastic thermodynamics \cite{mu6}, time-reversal operations \cite{mar2,mar1}, and the fluctuation theorem \cite{mg6,mg7,mg8}. The thermodynamic uncertainty relation (TUR) has recently emerged as a fundamental framework for estimating the entropy production from time-series data \cite{mg10,mg11,mg12,mg14,mg15}. Additionally, recent studies have extended entropy production analysis to non-Markovian systems \cite{mg9}.

   In our previous studies, we investigated the thermodynamic properties of Brownian heat engines operating in heat baths with temperature profiles that decrease quadratically \cite{mg40}, linearly, or in a piecewise constant manner \cite{mg41}. Using the Boltzmann-Gibbs nonequilibrium entropy framework and entropy balance equation, we analyzed key thermodynamic quantities such as entropy production and extraction rates by solving the system dynamics over time. Our findings highlight that the quadratic temperature profile yields the lowest efficiency but the highest particle velocity compared with the other arrangements. Understanding these thermodynamic relationships is crucial for predicting energy dissipation and optimizing the transport efficiency in biological systems.

	This study advances the understanding of nonequilibrium systems by deriving thermodynamic relations for a heat engine operating under an exponentially decreasing temperature profile. This arrangement offers valuable insights into the entropy production, energy dissipation, and transport dynamics.  Such an arrangement is  particularly relevant as it mimics the temperature distribution in local heating scenarios, such as systems heated by a laser beam, where heat decreases spatially. Our exact analytical solution reveal that the velocity, entropy production, and extraction rates were significantly higher than those in piecewise thermal arrangements. In addition, we analyz the thermodynamic behavior of Brownian particles moving through networks. The results show that thermodynamic quantities, such as velocity and entropy production rate, are independent of network size, while extensive quantities, such as total entropy, depend on the number of microstates. Systems with an exponentially decreasing temperature gradient exhibit much lower entropy production and extraction rates than those with linearly or quadratically varying profiles, indicating reduced irreversibility and complexity introduced by such thermal baths.

	We also show that  a Brownian particle in a ratchet potential with spatially varying temperature achieves directed motion even in the absence of  external forces. The thermal asymmetry and applied load generate a net particle current, offering insights for microscale and nanoscale transport systems in which controlled motion is crucial. Notably, adjusting the thermal arrangement alone enabled directed motion. Moreover,  by tuning the system parameters, such as barrier height, load, and noise intensity, the particle velocity can be optimized. This,  in turn, helps to  improve the efficiency of microfluidic devices and nanoscale sensors. We believe that these findings will deepen our understanding of thermally driven ratchets and the critical role of temperature asymmetry in Brownian motors.  Moreover, we explore how the efficiency and coefficient of performance (COP) depend on the motor parameters. Although these motors are generally less efficient than those operating under piecewise thermal arrangements, they achieve faster velocities. Conversely, they exhibit higher efficiency but lower velocities than motors functioning within linearly or quadratically decreasing thermal environments. Thus,  this study enhances the understanding of thermally driven ratchets, highlighting the role of temperature asymmetry in controlling the transport processes across various physical systems. Future work should focus on extending these findings to systems with many interacting particles, exploring collective dynamics, and validating theoretical predictions through experimental studies.

		Using the exact analytical results, we examine nonequilibrium thermodynamics as a general framework, with equilibrium thermodynamics as a limiting case. Equilibrium systems minimize the free energy or maximize the entropy, whereas nonequilibrium systems exhibit irreversible dynamics. This study addresses the following key questions: (1) Can a unified framework, similar to the Boltzmann distribution and Gibbs free energy, be developed for nonequilibrium systems? (2) How is entropy production, a hallmark of irreversibility, quantitatively linked to microscopic processes? (3) Why do some nonequilibrium systems maximize entropy production, while others follow different paths, and can a single principle explain this variability? By addressing these questions, we contribute to our perspective, providing insights into transitions between nonequilibrium and equilibrium, and tools to optimize energy-efficient systems in both natural and engineered environments.

		The remainder of this paper is organized as follows: Section II examines how the number of accessible states and entropy depend on the model parameters. Section III explores the behavior of thermodynamic rates, such as entropy production and extraction rates, as functions of model parameters. In Section IV, we analyze the velocity, coefficient of performance (COP) of the refrigerator, and efficiency. Section V investigates the temporal evolution of thermodynamic properties in a network of Brownian motors. Finally, Section VI presents the summary and conclusion.

\section{Effective Number of Accessible States and Entropy in Nonequilibrium Systems}

\subsection{Analysis of  Effective Number of Accessible States  in a Ratchet Potential with Exponential Temperature Profile}

  In this section, we consider  a Brownian motor moving along a ratchet potential coupled with an exponentially decreasing  thermal arrangement $T(x) = T_h e^{-\alpha x}$, where $\alpha = \frac{\ln\left(\frac{T_h}{T_c}\right)}{L}$, and periodic for $0 < x < L$. We show that   such  a thermal arrangement  leads to a  higher velocity, entropy production, and exaction rates compared to the piecewise constant thermal arrangement.  By considering discrete thermal and potential arrangements, we solve the master equation and provide an exact time-dependent solution, as detailed in Appendix I.  Detailed  derivations of thermodynamic quantities such as entropy, velocity, and entropy production are presented in  Appendix II. It should be noted that  such  localized temperature variations mimic  the temperature distribution of  laser-induced heating.

\subsection*{Effective Number of Accessible States}

Let us now explore the dependence on   the number of assessable states.   As an alternative approach, the effective number of accessible states, $\Omega_{\text{eff}}(t)$, helps explore  degree of irreversibility. For  a given  probability distribution $\{P_i(t)\}$,  we write the effective number of accessible states  as  
\begin{equation}
\Omega_{\text{eff}}(t) = \prod_{i=1}^N P_i(t)^{-P_i(t)}.
\end{equation}
Here,  $N$ denotes  the total number of microstates and $P_i(t)$ represents the probability that the system is in state $i$ at time $t$. The probabilities satisfy the following normalization condition: $\sum_{i=1}^N P_i(t) = 1, \quad 0 \leq P_i(t) \leq 1.$  Accordingly,  states with higher probabilities contribute more to $\Omega_{\text{eff}}(t)$, while less probable states contribute less. Consequently, $\Omega_{\text{eff}}(t)$  exhibits  an effective diversity of microstates. This approach corresponds to the geometric mean of a weighted probability distribution, which serves as an exponential measure based on the logarithmic contributions of $\{P_i(t)\}$.

In Appendix I,  we calculate the probability distributions  $P_i(t)$   in terms  of  the transition rates using the master equation framework. On the other hand, the Boltzmann entropy (using Eq. (1)) is given by
\begin{eqnarray}
S &=& k_B \ln \Omega_{\text{eff}}(t)\\ \nonumber 
 &=& k_B \left(-\sum_{i=1}^N P_i(t) \ln P_i(t)\right)
\end{eqnarray}, 
where $k_B$ is Boltzmann’s constant.  When the system is at equilibrium, all microstates are equally probable and as  a result one finds  $P_i = \frac{1}{N}, \quad \forall i.$ For a uniform distribution, $\Omega_{\text{eff}}(t) = N$ which  clearly indicates  the principle of entropy maximization for equilibrium thermodynamics.  Entropy maximization states that systems evolve toward states of maximum entropy, where all accessible microstates are equally probable. At equilibrium, this principle aligns with the minimization of the free energy, providing a consistent framework for isolated and closed systems.   For our model system (for equilibrium case) in the absence of an external load ($f=0$), potential ($E=0$), and with a uniform temperature ($T_h \to T_c$), we find that  
\begin{equation} 
\Omega_{\text{eff}}(t) = 3 \left(1 - e^{-\frac{3t}{2}}\right)^{\frac{2}{3} \left(-1 + e^{-\frac{3t}{2}}\right)}  \left(1 + 2 e^{-\frac{3t}{2}}\right)^{-\frac{1}{3} - \frac{2}{3} e^{-\frac{3t}{2}}}. \end{equation}
 In the limit $t \to \infty$   , this approaches $\Omega_{\text{eff}} = 3$, which represents the maximum entropy. This result is consistent because our   ratchet potential is consistent with that of the three lattice sites.

Our analysis   indicates that far from equilibrium, the effective number of accessible states $\Omega_{\text{eff}}(t)$ is generally less than the total number of microstates $N$. These results  ensure that $\Omega_{\text{eff}}(t) < N$ under nonequilibrium conditions, with $\Omega_{\text{eff}}(t)$ approaching  $N$ only at equilibrium. As illustrated in Fig. 1, for systems driven out of equilibrium $\Omega_{\text{eff}} < 3$. Notably, $\Omega_{\text{eff}}(t)$ is significantly larger for an exponential thermal arrangement than for a piecewise constant temperature profile. This indicates  that  systems  with exponential thermal arrangements are more irreversible than those  with piecewise constant temperature profiles.  
\begin{figure}[ht] 
\centering {     \includegraphics[width=6cm]{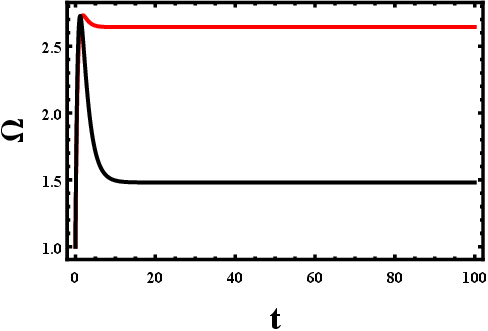} }
 \caption{ (Color online) The  $\Omega$  as a function of  $t$ and $\tau=6$   evaluated analytically via Eq. (9) for a given  $\epsilon=2.0$ and  $f=0.5$.  The topmost curve corresponds to an exponential temperature profile, while the bottom curve represents a piecewise constant temperature arrangement. This reflects  that  system  with exponential thermal arrangement is more irreversible than system  with a piecewise constant temperature profile. }   
\label{fig:sub} 
 \end{figure} 

From now on whenever  plot the  figures we use  rescaled load $\lambda = f/T_c$, barrier height $\epsilon = E/T_c$, and dimensionless temperature $\tau = T_h/T_c$. The quantity $\Omega$, as a function of $\epsilon$ and $\lambda$, is analytically evaluated using Eq.~(1) for $t = 100000.0$ and $\tau = 2.0$ in Fig. 2. The uppermost curve corresponds to an exponential temperature profile, while the lowermost curve represents a piecewise constant temperature configuration. These results demonstrate that systems with an exponential thermal arrangement exhibit a higher degree of irreversibility compared to those with a piecewise constant temperature profile. 
\begin{figure}[ht] 
\centering {     \includegraphics[width=6cm]{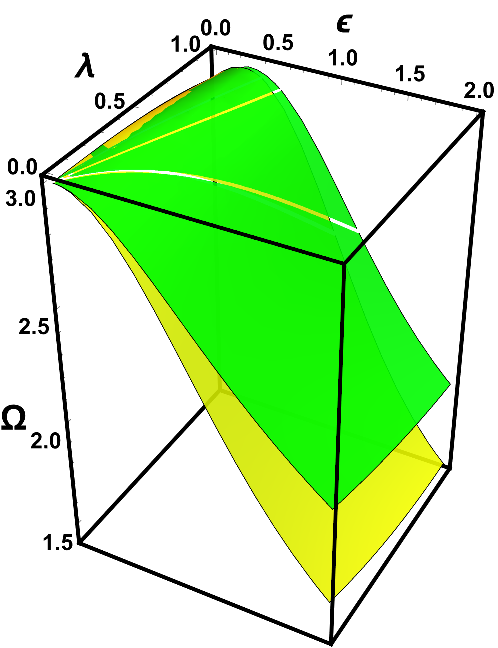} }
 \caption{ (Color online) The  $\Omega$  as a function of  $\epsilon$ and $\lambda$   evaluated analytically via Eq. (1) for a given  $t=100000.0$ and  $\tau=2.0$. The topmost curve corresponds to an exponential temperature profile, while the bottom curve represents a piecewise constant temperature arrangement. This reflects  that  system  with exponential thermal arrangement is more irreversible than system  with a piecewise constant temperature profile. }   
\label{fig:sub} 
 \end{figure} 
   
\subsection{Entropy and free energy in Nonequilibrium Systems} 

 We derive the probability distribution as a function of time, load $f$, and ratchet potential $E$  for  an exponentially decreasing temperature gradient that varies from  $T_h$ to $T_c$ (see Appendix I). Since the expressions for the probability distributions in the nonequilibrium case are lengthy, we do not present them here. Using these distributions, we calculate key thermodynamic quantities, including the entropy and free energy.

To gain insight into the irreversibility of these processes, we focus on studying entropy production. Using the exact analytical results, we address the critical question of how entropy production can be quantitatively linked to microscopic processes. By employing Gibbs entropy and solving the probability  distribution of microstates, we bridge microscopic dynamics with macroscopic thermodynamic properties. Notably, for systems governed by a master equation, the Boltzmann-Gibbs nonequilibrium entropy and entropy balance equation offer a reliable and robust framework. The Gibbs entropy is given by 
\begin{equation} S(t) = -k_B \sum_i P_i(t) \ln P_i(t), 
\end{equation} where $P_i(t)$ denotes the probability that the system is in state $i$ at time $t$.  This approach allowed us to compute macroscopic quantities (such as entropy production and entropy flow)  directly from microscopic probabilities. Theoretical studies \cite{mu1,mu2,mu3} have shown that nonequilibrium systems continuously produce and extract entropy. At steady state, the entropy  production rate  balances the  entropy extraction rate.  Thus, the dynamics described by    Gibbs entropy directly link microscopic stochastic processes  with  those of  macroscopic dynamics,  which in turn  improves  our understanding of irreversibility in non-equilibrium systems.

We now compare the entropy of the exponentially decreasing temperature case with piecewise constant thermal arrangements. Figure 3 shows entropy $S(t)$ as a function of time $t$ and $\lambda$  for fixed  values of $\tau = 2.0$ and  $\epsilon = 2.0$. The results clearly  show  that the entropy is significantly higher for an exponentially decreasing temperature gradient than for piecewise constant thermal arrangements.  
 \begin{figure}[ht]
\centering
{
    \includegraphics[width=6cm]{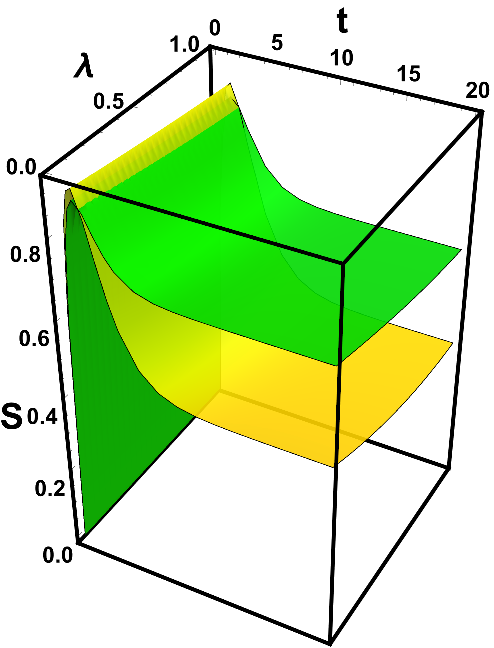}
}
\caption{ (Color online)   Entropy $S(t)$ as a function of time $t$ and $\lambda$  for a fixed  $\tau = 2.0$ and  $\epsilon = 2.0$. The top curve represents an exponential temperature profile, while the bottom curve corresponds to a piecewise constant arrangement. The results show that the system with the exponential thermal arrangement has considerably large entropy. 
} 
\label{fig:sub} 
\end{figure} 
  In equilibrium systems, principles such as the Gibbs free energy provide a well-established framework. Recent studies have suggested that Gibbs entropy also applies to nonequilibrium systems, offering a strong foundation for advancing the understanding of nonequilibrium thermodynamics.

 Figure 4 shows the free energy as a function of time $t$ plotted for $\tau = 6$, $\epsilon = 2.0$, and $f = 0.5$. The top curve represents the exponential thermal arrangement, whereas the bottom curve corresponds to the piecewise constant-temperature case.
\begin{figure}[ht]
\centering
{
    \includegraphics[width=6cm]{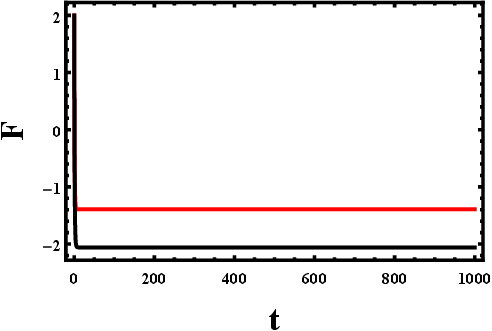}
}
\caption{  (Color online)  The free energy as a function of $t$, with $\tau = 6$, is plotted  for a fixed  $\epsilon = 2.0$ and $f = 0.5$. The top curve represents the exponential temperature case, whereas the bottom curve corresponds to the piecewise constant-temperature case.
} 
\label{fig:sub} 
\end{figure}

To gain  a better  understanding of the system, we analyze specific limiting cases. First, we consider the isothermal case, where $T_h \to T_c$ and $f  \to 0$. Under steady-state conditions, entropy $S$ can be expanded to small barrier heights $E$ as
\begin{equation}
S = \ln[3] - \frac{E^2}{3 T_c^2} + \frac{E^4}{12 T_c^4}.
\end{equation}
From this expression, it is clear that as $E \to 0$, the system approaches equilibrium and the entropy reduces to $S = \ln[3]$, which is the expected value at equilibrium. In the absence of ratchet potential, the system achieves maximum entropy. Conversely, for a small but nonzero $E$, the system deviates slightly from equilibrium, with entropy corrections proportional to $E^2$ and higher-order terms.

   Next, we consider the isothermal case with a nonzero force, where $E = 0$ and the system is driven out of equilibrium by an external load, $f$. Under steady-state conditions, expanding $S$ to a small $f$ yields
\begin{equation}
S = \ln[3] - \frac{f^4}{27 T_c^4} + \frac{f^5}{81 T_c^5}.
\end{equation}
In this case, when $f \to 0$, the entropy decreases to $S = \ln[3]$, corresponding to the equilibrium. However, for any nonzero $f$, the system is driven out of equilibrium, resulting in an increase in the entropy. This demonstrates that an external force alone, even in the absence of a potential, can generate non-equilibrium dynamics. These two limiting cases highlight the role of the potential barrier and external force in driving the system out of non-equilibrium states. While the system approaches equilibrium when $E$ or $f$ tends to zero, small expansions in these parameters reveal how they influence the entropy dynamics and deviations from equilibrium. 

 We emphasize that the principle of entropy maximization is fundamental to the equilibrium thermodynamics. This principle  states that systems evolve toward maximum entropy, where all microstates are equally probable. At equilibrium, this is directly linked  to free-energy minimization, forming a consistent framework for isolated and closed systems. In contrast, non-equilibrium systems display more complex behaviors, with entropy production influenced by external forces, thermal gradients, and nonlinearities. As nonequilibrium thermodynamics extends beyond equilibrium conditions, entropy maximization is only a special case. As shown earlier,  entropy maximizes or free energy is minimized only under a uniform probability distribution, $P_i(t) = 1/N$ for all $i$, a condition that applies exclusively at equilibrium.

This can be appreciated when  one writes the  probabilities as a function of time  for  the equilibrium cases for the isothermal case.  In the absence of a potential barrier ($E = 0$), load ($f = d = 0$) and when  ($t \to \infty$), we get the free energy 
 \begin{equation} F = -T_c \ln(3) 
\end{equation} 
which is the free energy of the three-state system  presented in this study.

This result can be verified using a statistical mechanical approach. In the limit $t \to \infty$, the probability distributions (shown in Appendix I) converge 
\begin{eqnarray}
P_1 &=& \frac{1}{1 + e^{-E/T_c} + e^{-2E/T_c}},\nonumber \\
P_2 &=& \frac{e^{-E/T_c}}{1 + e^{-E/T_c} + e^{-2E/T_c}},\nonumber \\
P_3& =& \frac{e^{-2E/T_c}}{1 + e^{-E/T_c} + e^{-2E/T_c}}.
\end{eqnarray}
The partition function is given 
\begin{equation}
Z = \sum_{i=1}^3 e^{-E_i/T_c} = 1 + e^{-E/T_c} + e^{-2E/T_c}.
\end{equation}
Using $Z$, the free energy and entropy are calculated as:
\begin{equation}
F = -T_c \ln(Z), \quad S = \ln(Z) + \bar{E}\beta,
\end{equation}
where $\bar{E}$ represents the average energy and $\beta = 1/T_c$.
In the limit $E \to 0$, the partition function simplifies to $Z = 3$. Consequently, the free energy converges to:
\begin{equation}
F = -T_c \ln(3),
\end{equation}
and the entropy approaches:
\begin{equation}
S = \ln(3).
\end{equation}
These results confirm the consistency of the thermodynamic relations under the isothermal case at equilibrium.

Equivalently, we can directly evaluate entropy $S$ in the absence of a load, $T_h=T_c$ and $E=0$  as 
\begin{widetext} 
\begin{eqnarray}
 S = \frac{1}{3} e^{-3t/2} \left[     -2 \left(-1 + e^{3t/2}\right) \ln\left(\frac{1}{3} - \frac{1}{3} e^{-3t/2}\right)      - \left(2 + e^{3t/2}\right) \ln\left(\frac{1}{3} + \frac{2}{3} e^{-3t/2}\right) \right]. 
\end{eqnarray} 
\end{widetext} 
In the limit $t \to \infty$, the entropy is simplified to \begin{equation} S = \ln(3), \end{equation} as expected, since the system has three lattice points.

The above analysis   indicates that nonequilibrium thermodynamics encompass a broader range of conditions than equilibrium; thus,  entropy maximization arises only as a limiting case. {\it  Thus, it  is my perspective that Gibbs entropy provides a universal framework linking microscopic dynamics to macroscopic thermodynamic properties, applicable to both equilibrium and non-equilibrium systems. In non-equilibrium thermodynamics, which encompasses systems with varying thermodynamic fluxes, thermal gradients, and external forces, entropy maximization emerges only as a special limiting case. Systems that are far from equilibrium require individual treatment based on their configuration, as they exhibit unique behaviors. This process begins by determining the probability distribution, which forms the foundation for analyzing the thermodynamic properties and understanding the interplay between irreversibility, energy dissipation, and entropy production. } 

\section{Entropy production and extraction  rates}

In this section,   we explore the dependence of  several thermodynamic relations, such as  entropy production and extraction rates, on the model parameters.   The  differentiation of the entropy with respect to time   leads to  
\begin{equation} 
\frac{dS_{\text{tot}}}{dt} = \dot{e}_p - \dot{h}_d.
 \end{equation} 
 Here, $(\dot{e}_p)$ denotes  the entropy production rate and $\dot{h}_d$ denotes the entropy extraction  rate. \(S_{\text{tot}}\) is the total entropy of the system. Since the expressions   for probability distribution as well as the transition rates are   analytically obtained,  we can write     the first and second laws of thermodynamics in terms of model parameters.  This, in turn, enables us  to uncover  several  new thermodynamic relations specific to these temperature profiles.

In general, the entropy production and extraction rates are substantially higher for an exponentially decreasing temperature profile than for a piecewise constant thermal arrangement. In Fig. 5, the entropy production rate $(\dot{e}_p)$ is plotted as a function of $\tau$ and $\epsilon$ for a given $\lambda = 0.1$. The figure depicts that the exponentially decreasing temperature profile exhibits significantly higher entropy, entropy production rate, and velocity than the piecewise constant thermal arrangement. 
\begin{figure}[ht]
\centering
{
    \includegraphics[width=6cm]{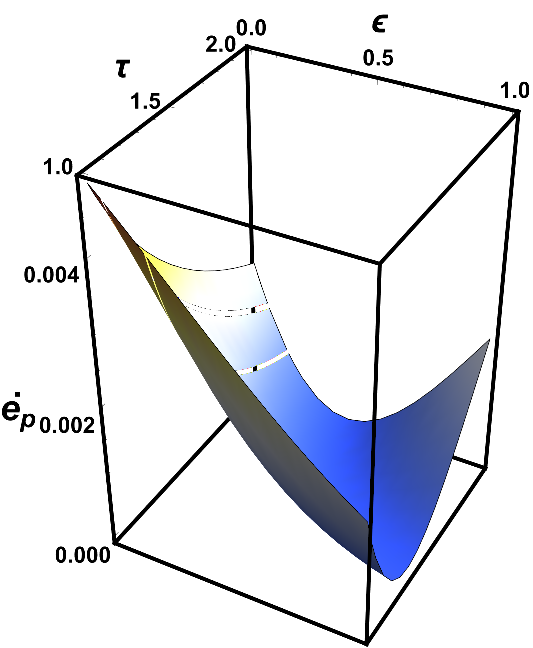}
}
\caption{ (Color online) Entropy production rate  $(\dot{e}_p)$  as a function of $\tau$ and $\epsilon$ for a given $\lambda = 0.1$.   } 
\label{fig:sub} 
\end{figure} 
Moreover, in Fig. 6, we plot the entropy production rate as a function of the time. As time progresses, the entropy production rate decreases and approaches a constant value. The figure also shows that the entropy production is significantly higher for the linearly decreasing temperature profile than for the piecewise constant case.  In Fig. 7, we plot the entropy extraction rate as a function of time, which again demonstrates that the extraction rate is higher for the exponentially decreasing temperature case.  Finally, in Fig. 8, we present ${\dot S}$ as a function of the time.
\begin{figure}[ht]
\centering
\includegraphics[width=6cm]{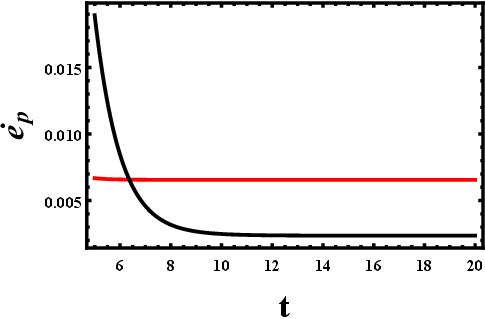}
\caption{(Color online) Entropy production rate as a function of $t$ for $\tau=6$, $\epsilon=2.0$ and $f=0.5$.  The top curve corresponds to an exponential temperature profile, whereas the bottom curve represents a piecewise constant-temperature arrangement.}
\label{fig:entropy_production}
\end{figure}
\begin{figure}[ht]
\centering
\includegraphics[width=6cm]{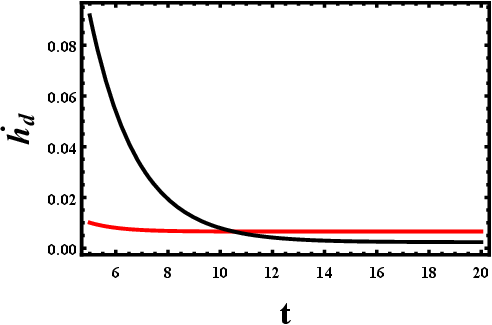}
\caption{(Color online) The entropy extraction rate is plotted as a function of $t$ for $\tau=6$, $\epsilon=2.0$, and $f=0.5$. The top curve represents an exponential temperature profile, while the bottom curve corresponds to a piecewise constant temperature arrangement.}
\label{fig:entropy_extraction}
\end{figure}
\begin{figure}[ht]
\centering
\includegraphics[width=6cm]{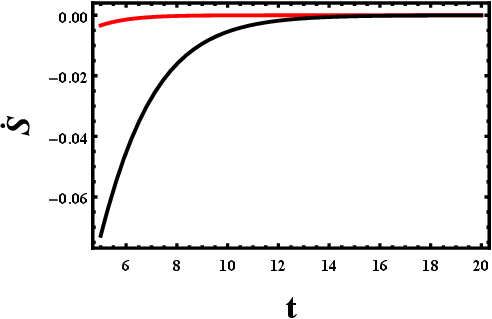}
\caption{(Color online) The change in entropy is plotted as a function of $t$ for $\tau=6$, $\epsilon=2.0$, and $f=0.5$. The top curve corresponds to a piecewise constant temperature arrangement, while the bottom curve represents an exponential temperature profile.
}
\label{fig:free_energy}
\end{figure}

Since the expressions for key thermodynamic relations are lengthy, we do not present them here. However, to provide basic insights, we explore some limiting cases. As the first limiting case, we consider the steady state in the absence of a ratchet potential, where the system remains out of equilibrium due to the external force. Under these conditions, the entropy production rate and heat dissipation rate are given by:
\begin{equation}
\dot{e}_p=\dot{h}_d=\frac{3 (-1 + e^{f/T_c}) f}{2 (T_c + 2 e^{f/T_c} T_c)}
\end{equation}
regardless of the thermal arrangement. As shown, both $\dot{e}_p$ and $\dot{h}_d$ are positive, indicating irreversible processes. In the limit $f \to 0$, we obtain $\dot{e}_p = \dot{h}_d = 0$, signifying that the system approaches equilibrium.

In the absence load, isothermal case and in the limit$E \to 0$,  for all thermal arrangement, we get 
\begin{equation}
\dot{e}_p=e^{-\frac{3t}{2}} \left(-\log\left[-1 + e^{\frac{3t}{2}}\right] + \log\left[2 + e^{\frac{3t}{2}}\right]\right)
\end{equation}
while integrating $\int_{0}^{\infty}(\dot{e}_p) dt$
leads to  $
S=e_p=\ln[3].
$
In an equilibrium system, entropy is determined by the number of accessible microstates. For our  three-state system in equilibrium, where each state has an equal probability, the entropy is given by $ S  = k_B \ln 3$
This represents the maximum entropy for a three-state system at equilibrium, where no external forces or temperature gradients disturb the system. In our case, we consider a three-state system operating under nonequilibrium conditions, where deviations from equilibrium modify the entropy due to external influences such as forces or thermal gradients.
Moreover \begin{equation}
\dot{E}_p=T_ce^{-\frac{3t}{2}} \left(-\log\left[-1 + e^{\frac{3t}{2}}\right] + \log\left[2 + e^{\frac{3t}{2}}\right]\right)
\end{equation}
while integrating $\int_{0}^{\infty}(\dot{E}_p) dt$
leads to  $
S^T={E}_p=T_c\ln[3].
$

\section{Velocity and  efficiency }

Brownian motors with ratchet potentials and spatially varying temperatures can achieve directed motion even without external forces. The thermal asymmetry and applied load generate a net particle current, offering valuable insights for microscale and nanoscale transport systems in which controlled motion is essential. Notably, adjusting the thermal arrangement alone enables directed motion.  By tuning the system parameters, such as barrier height, load, and noise intensity, the particle velocity can be optimized.  This, in turn,  improves the efficiency of microfluidic devices and nanoscale sensors. 

For exponentially decreasing temperature case,  after some algebra  the velocity is given as  
\begin{equation}
V=\frac{3 \left( e^{\frac{2 E}{T_c}} - e^{\frac{E \, T_c + f \, T_c + f \, T_h + (E + f) \, T_c \, \left(\frac{T_h}{T_c}\right)^{1/3}}{T_c \, T_h}} \right)}
{2 \left(1 + 2 \, e^{\frac{(E + f) \, \left(\frac{T_h}{T_c}\right)^{1/3}}{T_h}} \right) 
\left(e^{\frac{2 E}{T_c}} + e^{\frac{E + f}{T_h}} 
\left(e^{\frac{2 E}{T_c}} + e^{\frac{f}{T_c}}\right)\right)}.
\end{equation}
The stall force  at which current becomes is given as 
\begin{equation}
f' = \frac{-E \, T_c + 2 \, E \, T_h - E \, T_c \, \left(\frac{T_h}{T_c}\right)^{1/3}}
{T_c + T_h + T_c \, \left(\frac{T_h}{T_c}\right)^{1/3}}
\end{equation}

	The system sustains a nonzero velocity only in the presence of a symmetry-breaking field, such as a non-uniform temperature or an external load. To demonstrate this, we expand Eq. (19) for a small force in the absence of a ratchet potential and at uniform temperature. After some algebra, we obtain
		\begin{equation}
V = -\frac{f}{2 T_c} + \frac{f^2}{12 T_c^2}.
\end{equation}
Clearly, only in the limit $f \to 0$ does the velocity vanish, indicating that the external force is responsible for the unidirectional motion of the particle.  

In the absence of load $f$, the particle still attains unidirectional motion, but only in the presence of a ratchet potential and non-uniform thermal arrangement. Expanding Eq.~(19) for a small barrier height $E$, we obtain:
\begin{equation}
V = \frac{1}{6} \left(\frac{2}{T_c} - \frac{1 + (T_h/T_c)^{1/3}}{T_h}\right) E^2.
\end{equation}
In the limit $T_h \to T_c$, the velocity approaches zero, confirming that a unidirectional current arises owing to the thermal asymmetry and ratchet potential.

 Let us now  plot the velocity  as a function of  the key parameters.  The  velocity, as a function of the model parameters, exhibits similar patterns to those observed in systems with quadratic \cite{mg40}, linear, and piecewise constant \cite{mg41} thermal arrangements. As shown in Fig. 9, for a large load, the current reverses direction.  To compare the velocity between the exponential and piecewise constant-temperature cases, we computed their velocity ratio. Figure 10 shows that the velocity is significantly higher for the exponentially decreasing temperature case.
\begin{figure}[ht]
\centering
{
    \includegraphics[width=6cm]{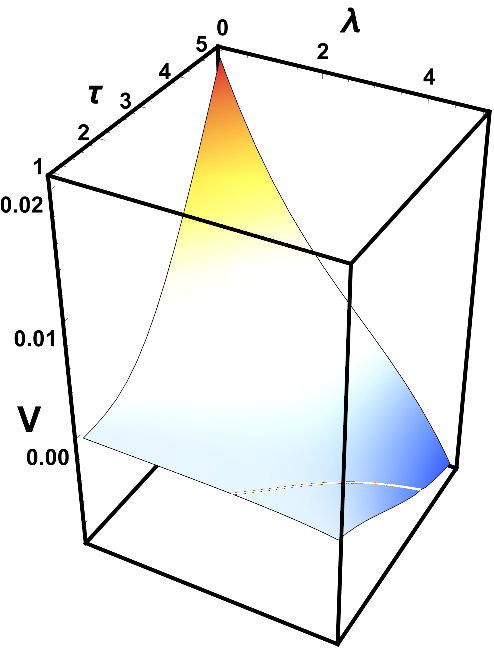}
}
\caption{ (Color online) The velocity as a function of  $\tau$ and $\lambda$   evaluated analytically  for a given   $\epsilon=6.0$.  } 
\label{fig:sub} 
\end{figure} 

\begin{figure}[ht]
\centering
{
    \includegraphics[width=6cm]{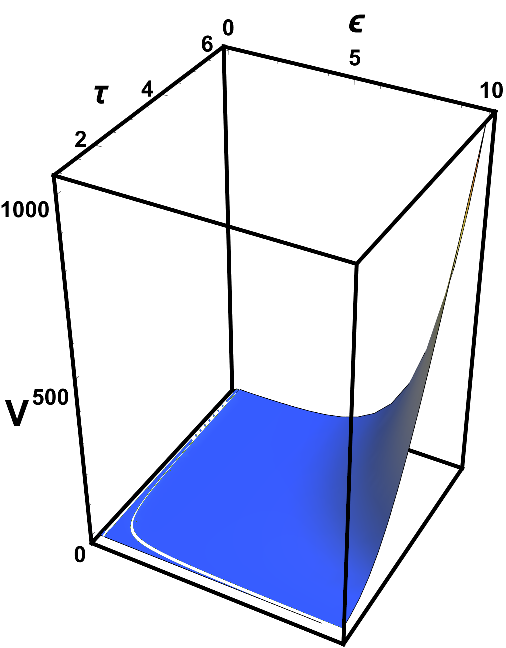}
}
\caption{ (Color online) The ratio of the velocity  between the exponential decreasing and piecewise constant temperature cases is evaluated analytically as a function of $\tau$ and $\epsilon$ for a given $\lambda = 0.1$.  The results highlight that the  efficiency  for  the exponential temperature case is considerably larger than that for the piecewise constant-temperature case. } 
\label{fig:sub} 
\end{figure}

The motor functions as a heat engine when the external force $f$ is less than the stall force $f'$. To analyze its performance, we examine the efficiency $\eta$. The efficiency of a Brownian motor is defined as the ratio of useful output work, given by $W=3f$, to the total energy input, $E_{\text{in}}=2E+2f$. For a system driven by a ratchet potential and a thermal gradient, the efficiency in the quasistatic limit is given by:
\begin{equation}
\eta = 1 - \frac{T_c \left(1 + \left(\frac{T_h}{T_c}\right)^{1/3}\right)}{2 T_h}.
\end{equation}
Here, $T_h$ and $T_c$ represent the temperatures of the hot and cold reservoirs, respectively. This efficiency is significantly lower than the Carnot efficiency, which represents the theoretical maximum efficiency in the quasistatic limit for a piecewise constant-temperature system.

A key observation is that Brownian motors operating under an exponentially decreasing temperature profile generally achieve higher velocities but at the cost of lower efficiency compared to piecewise constant thermal arrangements. In contrast, motors operating under linearly or quadratically varying temperature profiles exhibit very high velocities but lower efficiency, reflecting the trade-off between speed and energy conversion efficiency.

When $f > f'$, the motor transitions to refrigeration mode, where it pumps heat against the thermal gradient. The coefficient of performance (COP) is defined as the ratio of heat extracted from the cold reservoir, $Q_{\text{cold}} = 2E - f$, to the work input, $W_{\text{input}} = 3f$. In the quasistatic limit, the COP is given by:
\begin{equation}
COP=-\frac{T_c \left(1 + \left(\frac{T_h}{T_c}\right)^{1/3}\right)}
{T_c - 2 T_h + T_c \left(\frac{T_h}{T_c}\right)^{1/3}}
\end{equation}
\begin{figure}[ht]
\centering
{
    \includegraphics[width=6cm]{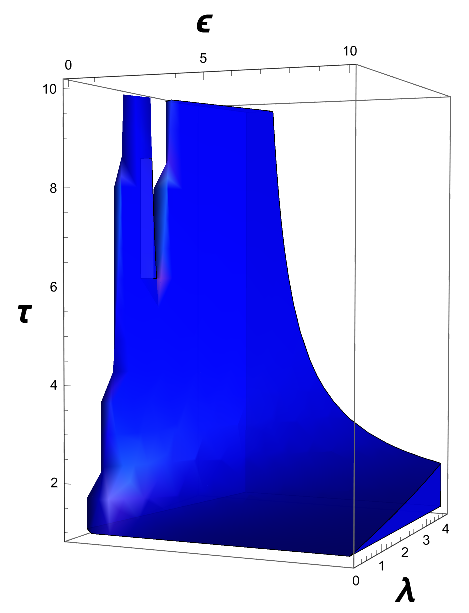}
}
\caption{ (Color online) The phase diagram illustrates the regime where the engine operates as a refrigerator, plotted as a function of $\epsilon$, $\lambda$, and $\tau$. This diagram provides insight into the parameter space in which the motor transitions from heat engine mode to refrigeration mode, highlighting the dependence of performance on system characteristics.
. } 
\label{fig:sub} 
\end{figure} 

The phase diagram shown in Fig.11 illustrates the regime where the engine operates as a refrigerator, plotted as a function of $\epsilon$, $\lambda$, and $\tau$. This diagram provides insight into the parameter space in which the motor transitions from heat engine mode to refrigeration mode, highlighting the dependence of performance on system characteristics.

Our analysis reveals that Brownian motors operating under an exponentially decreasing temperature gradient generally exhibit a lower COP compared to those in a piecewise constant thermal environment. This is because a continuously varying temperature gradient leads to additional entropy production, increasing energy dissipation. However, these motors still offer advantages in terms of enhanced velocity, making them suitable for applications in nanoscale energy harvesting and transport.

\section{Brownian Particle Dynamics in a Network of $M$ Ratchet Potentials}

In this work let us consider  a Brownian  particle moving in a network.   We show that similar  to our previous study \cite{mg40}, we find that the rates of thermodynamic quantities such as velocity, entropy production rate, and entropy extraction rate are independent of network size at steady state. However, extensive properties like entropy, entropy production, and entropy extraction increase with network size, even in the steady state. We will then  compare  this  thermodynamic relations with  systems operating between hot and cold baths and those where the temperature decreases exponentially along the reaction coordinate.

{\it Master equation approach. \textemdash} One can directly  calculate the thermodynamic relations. For instance the dynamics of a Brownian particle moving within a network of two ratchet potentials (see Fig. 14) can be described by considering a periodic boundary condition as:
\begin{eqnarray}
{dp_{1}(t) \over dt } &=& P_{1,2}p_2 - P_{2,1}p_1 + P_{1,3}p_3 - P_{3,1}p_1, \\ \nonumber
{dp_{2}(t) \over dt } &=& P_{2,1}p_1 - P_{1,2}p_2 + P_{2,3}p_3 - P_{3,2}p_2, \\ \nonumber
{dp_{2}(t) \over dt } &=& P_{2,1}p_1 - P_{1,2}p_2 + P_{2,3}p_3 - P_{3,2}p_2, \\ \nonumber
{dp_{3}(t) \over dt } &=& P_{3,2}p_2 - P_{2,3}p_3 + P_{3,1}p_1 - P_{1,3}p_3, \\ \nonumber
{dp_{3}(t) \over dt } &=& P_{3,2}p_2 - P_{2,3}p_3 + P_{3,1}p_1 - P_{1,3}p_3.
\end{eqnarray}

The above  differential equations can be simplified  to 
\begin{eqnarray}
\frac{dp_1(t)}{dt} &=& P_{1,2}p_2 - P_{2,1}p_1 + P_{1,3}p_3 - P_{3,1}p_1, \\ \nonumber
\frac{dp_2(t)}{dt} &=& P_{2,1}p_1 - P_{1,2}p_2 + P_{2,3}p_3 - P_{3,2}p_2, \\ \nonumber
\frac{dp_3(t)}{dt} &=& P_{3,2}p_2 - P_{2,3}p_3 + P_{3,1}p_1 - P_{1,3}p_3.
\end{eqnarray}
Note that  even though  the above differential equations reduce to three  equations ,  entropy does not simplify in the same way. The total entropy depends on the number of sites in the network. For a network consisting of $N$ sites, the entropy $S$ is expressed as:
\begin{equation}
S = -k_B \sum_{i=1}^N P_i(t) \ln P_i(t),
\end{equation}
where $(P_i(t))$ represents the probability of the particle occupying state $i$ at time $t$. 
This dependency arises because entropy is an extensive quantity which depends on   with the number of microstates in the system. While the dynamical rates such as velocity, entropy production  rate, and entropy extraction rate are independent of network size, the total entropy grows proportionally with the number of sites, reflecting the increasing complexity of the system.  For instance, for any  network size $M$, at steady state  we find the velocity
\begin{equation}
V=\frac{3 \left( e^{\frac{2 E}{T_c}} - e^{\frac{E \, T_c + f \, T_c + f \, T_h + (E + f) \, T_c \, \left(\frac{T_h}{T_c}\right)^{1/3}}{T_c \, T_h}} \right)}
{2 \left(1 + 2 \, e^{\frac{(E + f) \, \left(\frac{T_h}{T_c}\right)^{1/3}}{T_h}} \right) 
\left(e^{\frac{2 E}{T_c}} + e^{\frac{E + f}{T_h}} 
\left(e^{\frac{2 E}{T_c}} + e^{\frac{f}{T_c}}\right)\right)}.
\end{equation}
The stall force  at which current becomes is given as 
\begin{equation}
f = \frac{-E \, T_c + 2 \, E \, T_h - E \, T_c \, \left(\frac{T_h}{T_c}\right)^{1/3}}
{T_c + T_h + T_c \, \left(\frac{T_h}{T_c}\right)^{1/3}}
\end{equation}

Since the rate equations are independent of the network size $M$, their expressions have already been discussed in the previous section. Here, we further investigate extensive quantities, such as entropy, as a function of the model parameters.  Since the probability for each lattice site is exactly obtained as a function of the model parameters, we can construct a complete picture of $S$. However, its expression is too complex to present here. Instead, to provide insight for readers, we explore some key limiting cases.
Let us now explore the entropy for $M=4$ (four Brownian particles arranged in a network, sharing the same endpoint with periodic boundary conditions). 

For the isothermal case (at steady state) and in the absence of a ratchet potential, expanding $S$ for small $f$ gives:
\begin{equation}
S =  \ln[9] - \frac{2 f^4}{81 T_c^4}.
\end{equation}
In the limit $f \to 0$, the entropy approaches $S \to \ln[9]$, as expected.

Similarly, in the absence of load (at steady state) and under isothermal conditions, expanding $S$ for small $E$ yields:
\begin{equation}
S =  \ln[9] - \frac{2 E^2}{9 T_c^2}.
\end{equation}
Once again, in the limit $E \to 0$, the entropy converges to $S \to \ln[9]$, confirming the expected equilibrium behavior.
Care must be taken here, as for $M=1$, the entropy is $S = \ln[3]$, for $M=2$, it is $S = \ln[5]$, and for $M=4$, it is $S = \ln[9]$. This arises because these repetitive network structures share the same nodes at both ends. As shown in Fig.~13, for $M=2$, the system consists of five lattice sites.

{\it Generating function approach. \textemdash} In this section, we demonstrate that the rate of thermodynamic relations remains independent of the network size $M$ at steady state, adapting the mathematical approach introduced by Goldhirsch \emph{et al.} \cite{mg24,mg25}. This methodology was also employed in our recent work to compute the motor velocity \cite{mg40}.

For clarity, we first review the underlying mathematical technique. Consider a segment with sites $0, 1, 2, \dots, N$ as illustrated in Fig. 1b. Let $P_{2,1}$ denote the probability per unit time step for the walker to transition from site $1$ to $2$, and $P_{1,2}$ the probability for the reverse transition, subject to the condition $P_{1,2} + P_{2,1} \leq 1$. The probability $P_{w}(n)^{+}$ represents the likelihood of the walker starting at $j=0$ and reaching $j=N$ in $n$ steps in the rightward direction.

The mean first passage time (MFPT) to reach $j=N$ is given by
\begin{equation}
\left\langle t ^+\right\rangle = \frac{\sum_{n=0}^{\infty} n P_{w}(n)^{+}}{\sum_{n=0}^{\infty} P_{w}(n)^{+}}.
\end{equation}

The corresponding generating function in terms of the parameter $\phi$ is expressed as:
\begin{equation}
G_{N}(\phi)^{+} = \sum_{n=0}^{\infty} e^{i\phi n} P_{w}(n)^{+},
\end{equation}
where $0 < \phi < 2\pi$. The MFPT can then be rewritten in terms of the generating function as:
\begin{equation}
t^+ = \left.\frac{d}{d (i\phi)} G_{N}(\phi)^{+} \right|_{\phi=0}.
\end{equation}

The probabilities $P_{1}, P_{2}, \dots$ can be represented in terms of $\phi$ as $P_{1} e^{i\phi}, P_{2} e^{i\phi}, \dots$. The probability that the walker remains at $j=0$ after $n$ steps is $(1-P_{1})^n$, and the generating function is given by:
\begin{eqnarray}
X_{1-P_{1,2}} &=& \sum_{n=0}^{\infty} (1-P_{1,2})^n e^{i\phi n} \\
&=& \frac{1}{1-(1-P_{1})e^{i\phi}}.
\end{eqnarray}

Similarly, the probability of remaining at site $j$ for $n$ steps is:
\begin{eqnarray}
X_{1-P_{1,2}-P_{2,1}} &=& \sum_{n=0}^{\infty} (1-P_{1,2}-P_{2,1})^n e^{i\phi n} \\
&=& \frac{1}{1-(1-P_{1,2}-P_{2,1})e^{i\phi}}.
\end{eqnarray}

The probability of reaching $j=N$ without returning to $j=0$ is defined by the generating function:
\begin{equation}
T_{N}^{+} = \sum_{n=0}^{\infty} e^{i\phi n} U(n)^{+},
\end{equation}
whereas the probability of returning to $j=0$ without reaching $j=N$ is given by:
\begin{equation}
Q_{N}(\phi)^{+} = \sum_{n=0}^{\infty} e^{i\phi n} V(n)^{+}.
\end{equation}

Through recursive iterations, the generating function for the first passage time is formulated as:
\begin{equation}
G_{N}(\phi)^{+} = \frac{X_{1-P_{1,2}} T_{N}^{+}}{1 - X_{1-P_{1,2}} Q_{N}^{+}(\phi)}.
\end{equation}

For a Brownian particle navigating a complex network, the MFPT generating function in the right direction is:
\begin{equation}
G_{M}(\phi)^{+} = \frac{X'_{1-P{1,2}}(T_{{1}}^{+}+\dots+T_{{M}}^{+})}{1 - X'_{1-P{1,2}}(Q_{{1}}^{+}+\dots+Q_{{M}}^{+})}.
\end{equation}
Since the networks exhibit repetitive behavior, the expression $X'_{1-P_{1,2}}$ can be written as $X_{1-P_{1,2}}/M$. Moreover, the summation of the terms $(T_{{1}}^{+}+\dots+T_{{M}}^{+})$ simplifies to $M T_{N}^{+}$, and similarly, $(Q_{{1}}^{+}+\dots+Q_{{M}})$ results in $M Q_{N}$. Consequently, Equations (27) and (28) are found to be exactly equivalent, which demonstrates that the velocity and other thermodynamic rates are independent of the lattice size $M$.
Following similar steps, the left-direction generating function takes the form:
\begin{equation}
G_{M}(\phi)^{-} = \frac{X'_{1-P{3,1}}(T_{{1}}^{-}+\dots+T_{{M}}^{-})}{1 - X'_{1-P{3,1}}(Q_{{1}}^{-}+\dots+Q_{{M}}^{-})}.
\end{equation}
Considering $M$ identical potentials, the net average velocity, independent of lattice size $M$, simplifies to:
\begin{equation}
V\approx\frac{3 \left( e^{\frac{2 E}{T_c}} - e^{\frac{E \, T_c + f \, T_c + f \, T_h + (E + f) \, T_c \, \left(\frac{T_h}{T_c}\right)^{1/3}}{T_c \, T_h}} \right)}
{2 \left(1 + 2 \, e^{\frac{(E + f) \, \left(\frac{T_h}{T_c}\right)^{1/3}}{T_h}} \right) 
\left(e^{\frac{2 E}{T_c}} + e^{\frac{E + f}{T_h}} 
\left(e^{\frac{2 E}{T_c}} + e^{\frac{f}{T_c}}\right)\right)}.
\end{equation}
Using the method of generating functions, we demonstrate that the rates, such as velocity, are independent of the number of lattice sites.

\section{Summary and Conclusion}

 In this study,  we  explore the dependence of several thermodynamic relationships    for a  Brownian particle that walks  in ratchet potential, coupled with an exponentially decreasing temperature profile. Such a thermal  arrangement   mimics  localized heating, in which  the temperature  decreases spatially. Our exact analytical solutions show that the velocity, entropy production, and extraction rates are significantly higher than those observed in piecewise thermal arrangements.

Moreover,  we also analyze   the thermodynamic behavior of Brownian particles in the networks. The analytical results indicate   that    the  relations for thermodynamic rates, such as velocity and entropy production rate, remain independent of the network size. However,  extensive quantities, such as the total entropy,  depend on the number of microstates. The analytical results also indicate that the system with an  exponentially decreasing temperature gradient exhibits lower entropy production and extraction rates than those with linearly or quadratically varying thermal cases.

Furthermore, we show that a Brownian particle taht walks  in a ratchet potential  coupled  with spatially varying temperature achieves directed motion even without external forces. The presence of symmetry-breaking  fields  such as thermal asymmetry  and external forces  generates unidirectional motion.  The magnitude and direction of the particle motion can be adjusted by   varying  system parameters such as barrier height, load, and noise intensity.  Thus  the study presented in the work helps to improve the efficiency of microfluidic devices and nanoscale sensors by  deepen our  understanding of thermally driven ratchets and the role of temperature asymmetry in Brownian motors.

We also investigate the relationship between the efficiency and coefficient of performance (COP) for different thermal arrangements. We show that  Brownian motors operating under an exponentially decreasing temperature gradient generally have a lower efficiency than those with piecewise constant thermal profiles. However,  these motors achieve higher velocities. In comparison to linearly or quadratically varying thermal arrangement cases, the exponential thermal arrangement  yields  a higher efficiency but reduced speed.

Via the  exact analytical results, we examine nonequilibrium thermodynamics as a general framework, with equilibrium thermodynamics emerging as a special limiting case. We show that equilibrium systems minimize the free energy or maximize the entropy, whereas nonequilibrium systems exhibit irreversible dynamics. This study addresses the following fundamental questions: (1) Can a unified framework, similar to the Boltzmann distribution and Gibbs free energy, be developed for non-equilibrium systems? (2) How is entropy production, a key measure of irreversibility, quantitatively linked to microscopic processes? (3) Why do some nonequilibrium systems maximize entropy production, while others follow different paths, and can a unifying principle explain this variability?

By addressing these questions, we provide insights into the transitions between nonequilibrium and equilibrium systems, offering tools to optimize energy-efficient processes in both natural and engineered environments. Future research should extend these findings to systems with many interacting particles, investigate the collective dynamics, and validate theoretical predictions through experimental studies. In conclusion, this  study advances the understanding of non-equilibrium thermodynamics by deriving thermodynamic relations for a heat engine operating under an exponentially decreasing temperature profile. This setup provides key insights into the entropy production, energy dissipation, and transport dynamics. Notably, such a thermal arrangement mimics real-world heating scenarios such as laser-induced thermal gradients.

\section*{ Appendix A1}

To gain insights into the behavior of a Brownian particle operating between hot and cold thermal baths, we derive its probability distribution. The thermodynamic properties of a system with an exponentially decreasing temperature profile are compared to those of a Brownian particle in a piecewise constant thermal arrangement:
\begin{eqnarray}
T_j &=& T_h, \quad \text{for site } j = 0, \\
T_j &=& T_c, \quad \text{for sites } j = 1 \text{ and } j = 2.
\end{eqnarray}
 For  the exponentially decreasing temperature profile case,  the temperature at site  $i$ can be expressed as:
\begin{equation}
T_i = T_h e^{-\alpha i},
\end{equation}
where $\alpha$ is given by:
\begin{equation}
\alpha = \frac{\ln\left(\frac{T_h}{T_c}\right)}{3}
\end{equation}
and $i$ runs from $i=0$ to $i=2$.
Here, $T_h$ and $T_c$ are the temperature of hot and cold baths, respectively. $x_i$ represents the position.

The dynamics of the particle moving from one lattice site to the next is assumed to occur randomly. The probability of this transition is governed by the energy barrier that the particle must overcome and the temperature of the thermal reservoir to which it is coupled. Specifically, the transition rate per unit time for the particle to jump from site $i$ to $i+1$ is given by:
\begin{equation}
\Gamma e^{-\frac{\Delta E}{T_i}},
\end{equation}
where $\Delta E = U_{i+1} - U_i$, and $\Gamma$ represents the attempt frequency of the particle per unit time. Here, Boltzmann's constant $k_B$ and $\Gamma$ are taken to be unity. During each jump attempt, the particle first determines the direction of movement—either left or right—with equal probability. The jump is then governed by the Metropolis algorithm \cite{mg40}. If $\Delta E \leq 0$, the jump occurs with certainty; otherwise, if $\Delta E > 0$, the jump occurs with probability:
\begin{equation}
\exp\left(-\frac{\Delta E}{T_i}\right).
\end{equation}
The dynamics of a Brownian particle moving within a network (in the presence of load $f$) can be described by the Master equation:
\begin{equation} 
\frac{dp_{n}}{dt}=\sum_{n\neq n'}\left(P_{nn'}p_{n'}-P_{n'n}p_{n}\right),\quad n,n'=1,2,3.
\end{equation}

For a single ratchet potential case $M=1$, considering a periodic boundary condition, one can express Eq. (50) as
\begin{eqnarray}
{dp_{1}(t) \over dt }&=& P_{1,2}p_2-P_{2,1}p_1 +P_{1,3}p_3-P_{3,1}p_1  \\ \nonumber
{dp_{2}(t) \over dt }&=& P_{2,1}p_1-P_{1,2}p_2+P_{2,3}p_3-P_{3,2}p_2   \\ \nonumber
{dp_{3}(t) \over dt }&=& P_{3,2}p_2-P_{2,3}p_3 +P_{3,1}p_1-P_{1,3}p_3.  
\end{eqnarray}
Here $P_{n'n}$ is given by the Metropolis rule. For example,$
P_{21} = \frac{1}{2} e^{-\frac{E+f}{t1}}={\mu a\over 2}$, $ P_{12} = \frac{1}{2}$, $ P_{32} = \frac{1}{2} e^{-\frac{E+f}{t2}}={\nu b\over 2}$, $ P_{23} = \frac{1}{2}$, $P_{13} = \frac{1}{2} e^{-\frac{2E-f}{t3}}={\mu^2 \over 2a}$, $ P_{31} = \frac{1}{2}$
where $
\mu = e^{-\frac{E}{t1}} $, $ \nu = e^{-\frac{E}{t2}}$,
$a = e^{-\frac{f }{t1}} $  and  $ b = e^{-\frac{f}{t2}}$. Here $t_1=T_h$, $t_2=T_i = T_h e^{-\alpha }$ and $t_3=T_c$.
 \begin{figure}[ht]
\centering
{
    \includegraphics[width=6cm]{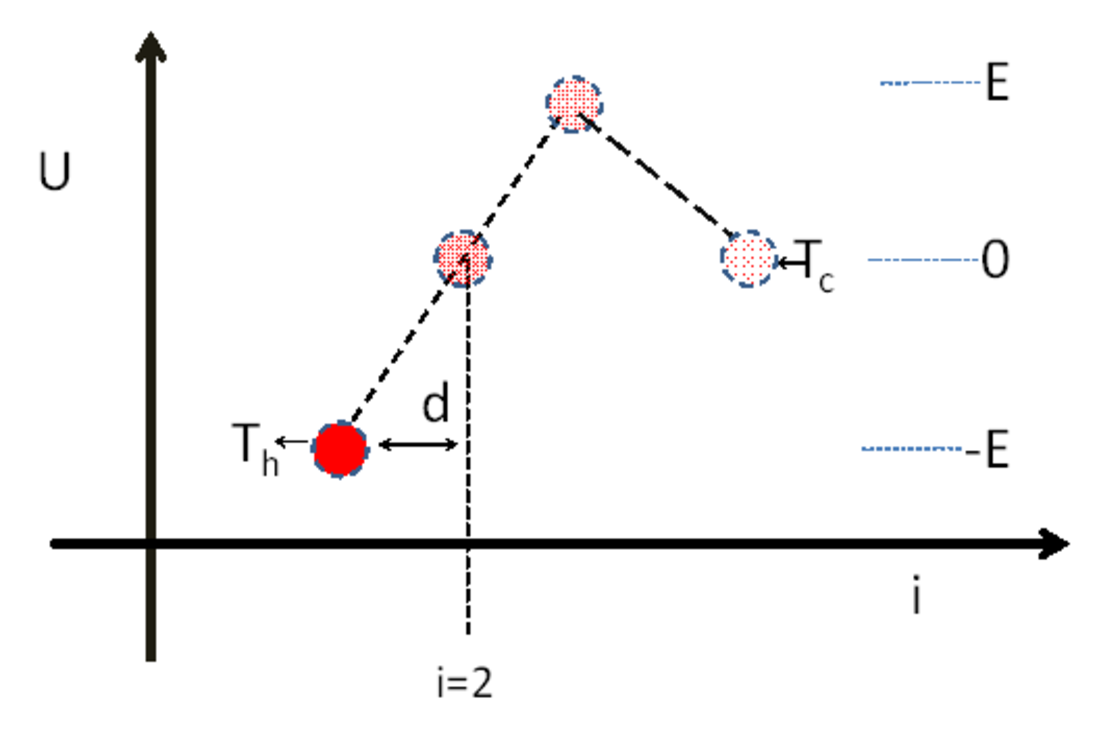}}
\caption{(Color online)  The schematic diagram for a Brownian particle that walks in a discrete ratchet potential coupled with a exponentially decreasing temperature heat bath.  The temperature for the heat bath decreases from $T_h$ to $T_{c}$ according to Eq. (46). } 
\end{figure}
The above  equation can be rewritten in the form
${d \vec{p}  \over dt}= {\bold P}\vec{p}$.  
${\bold P}$ has a form
\begin{equation} 
{\bold P}= \left( \begin{array}{ccc}
{-\mu_1 a_1\over 2}-{\mu_2^2\over 2a_2} & {1\over 2} & {1\over 2} \\
{\mu_1 a_1 \over 2} & {-1-\nu b\over 2} & {1\over 2}\\
{\mu_2^2 \over 2a_2} & {\nu b\over 2} & -1 \end{array} \right).
\end{equation}

For a particle that is initially  situated at  site $i=2$,  the time dependent  normalized probability distributions after solving Eq. (50) are given as  
\begin{eqnarray}
p_{1}(t)&=&c_{1}\frac{a (2+\nu b)}{\mu \left(\mu+\left(a^2+\mu\right) \nu b\right)}+\\ \nonumber
& &c_{2} e^{-\frac{\left(a+a^2 \mu+\mu^2\right) t}{2 a}} \left(-1+\frac{a
(-1+a \mu)}{-\mu^2+a \nu b}\right),\\
p_{2}(t)&=&-c_{3} e^{\frac{1}{2} t (-2-\nu b)}-c_{2}\frac{a\text{  }e^{-\frac{\left(a+a^2 \mu+\mu^2\right) t}{2 a}} (-1+a \mu)}{-\mu^2+a
\nu b}+\\ \nonumber
&&c_{1}\frac{ \left(2 a^2+\mu\right)}{\mu+\left(a^2+\mu\right) \nu b},\\
p_{3}(t)&=&c_{1}+c_{2} e^{-\frac{\left(a+a^2 \mu+\mu^2\right) t}{2 a}}+
c_{3} e^{\frac{1}{2} t (-2-\nu b)}
\end{eqnarray}
where 
\begin{eqnarray}
c_{1}&=& \frac{\mu \left(\mu+\left(a^2+\mu\right) \nu b\right)}{\left(a+a^2 \mu+\mu^2\right) (2+\nu b)},\\
c_{2}&=& -\frac{a}{\left(a+a^2 \mu+\mu^2\right) \left(-1+\frac{a (-1+a \mu)}{-\mu^2+a \nu b}\right)},\\
c_{3}&=& -\frac{\mu \left(\mu+a^2 \nu b+\mu \nu b\right)}{\left(a+a^2 \mu+\mu^2\right) (2+\nu b)}+ \\ \nonumber
&&\frac{a}{\left(a+a^2 \mu+\mu^2\right) \left(-1+\frac{a (-1+a \mu)}{-\mu^2+a\nu b}\right)}.
\end{eqnarray}

\begin{figure}[ht]
\centering
{
    \includegraphics[width=6cm]{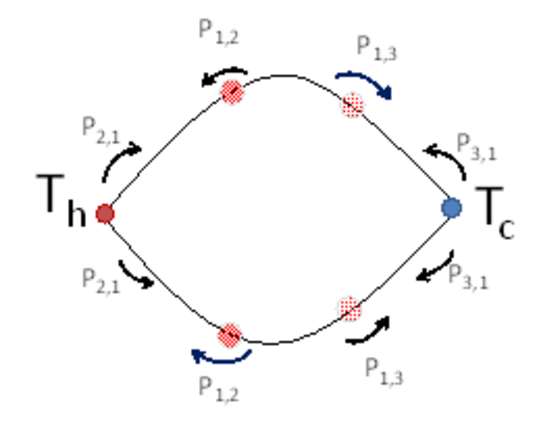}}
\caption{(Color online)  The top view that shows two thermal ratchets arranged in a lattice under periodic boundary conditions, where the temperature within each ratchet potential decreases quadratically from the hotter bath ($T_h$) to the colder bath ($T_c$). Both ratchet potentials share the same hot and cold baths.} 
\end{figure}

 Similarly, the dynamics of a Brownian particle moving within a network of two ratchet potentials $M=2$  (see Fig. 14) can be described by considering a periodic boundary condition as 

\begin{eqnarray}
{dp_{1}(t) \over dt }&=& P_{1,2}p_2-P_{2,1}p_1 +P_{1,3}p_3-P_{3,1}p_1  \\ \nonumber
{dp_{2}(t) \over dt }&=& P_{2,1}p_1-P_{1,2}p_2+P_{2,3}p_3-P_{3,2}p_2   \\ \nonumber
{dp_{2}(t) \over dt }&=& P_{2,1}p_1-P_{1,2}p_2+P_{2,3}p_3-P_{3,2}p_2   \\ \nonumber
{dp_{3}(t) \over dt }&=& P_{3,2}p_2-P_{2,3}p_3 +P_{3,1}p_1-P_{1,3}p_3  \\ \nonumber
{dp_{3}(t) \over dt }&=& P_{3,2}p_2-P_{2,3}p_3 +P_{3,1}p_1-P_{1,3}p_3.
\end{eqnarray}
Here
\begin{eqnarray}
P_{21}&=&{1 \over 2}e^{-(E+f)/T_{1}}, ~P_{12}={1 \over 2}\nonumber \\
P_{32}&=&{1 \over 2}e^{-(E+f)/T_{2}},P_{23}={1 \over 2}\nonumber \\
P_{13}&=&{1 \over 2},P_{31}={1 \over 2} e^{-(2E-f)/T_{3}}. 
\end{eqnarray}

The above  equation can be rewritten in the form
${d \vec{p}  \over dt}= {\bold P}\vec{p}$.  
${\bold P}$ has a form 
\begin{eqnarray} 
{\bold P}= \left( \begin{array}{ccccc}
{-2\mu_1 a_1\over 2}-{2\mu_2^2\over 2a_2} & {1\over 2} & {1\over 2} &{1\over 2} & {1\over 2} \\
{\mu_1 a_1 \over 2} & {-1-\nu b\over 2} & 0&{1\over 2}&0\\
{\mu_1 a_1 \over 2}  & 0&{{-1-\nu b\over 2}} & 0&{1\over 2}\\
{\mu_2^2 \over 2a_2} &{\nu b\over 2} &0&-1& 0\\
{\mu_2^2 \over 2a_2}&0&{\nu b\over 2}&0&-1\end{array} \right)
\end{eqnarray} 
where
$\mu_1=e^{-E/t_1}$, $\nu=e^{-(E+f)/2_2}$,  $a_1 = e^{-f / t_1}$,  $\mu_2=e^{-E/t_3}$ and $a_2 = e^{-f / t_3}$.

The expressions for $p_{1}(t)$, $p_{2}(t)$,  $p'_{2}(t)$, $p_{3}(t)$ and  $p'_{3}(t)$ for the particle which is initially  situated at  site $i=2$ is given as 
\begin{widetext}
\begin{eqnarray}
p_1 &=& \frac{a_2 (2 + \nu b)}{a_1 a_2 \mu_1 \nu b + \mu_2^2 (1 + \nu b)} c_1 
+ \left( -2 + \frac{2 a_2 (-1 + 2 a_1 \mu_1)}{-2 \mu_2^2 + a_2 \nu b} \right) 
\exp\left[ t \left( -\frac{1}{2} - a_1 \mu_1 - \frac{\mu_2^2}{a_2} \right) \right] c_3 \\
p_2 &=& \frac{2 a_1 a_2 \mu_1 + \mu_2^2}{a_1 a_2 \mu_1 \nu b + \mu_2^2 (1 + \nu b)} c_1 
- \frac{1}{\nu b} \exp\left[ -\frac{1}{2} t \right] c_2 
+ \left( \frac{a_2 - 2 a_1 a_2 \mu_1}{-2 \mu_2^2 + a_2 \nu b} \right) 
\exp\left[ t \left( -\frac{1}{2} - a_1 \mu_1 - \frac{\mu_2^2}{a_2} \right) \right] c_3 \nonumber \\
&& - \exp\left[ t \left( -1 - \frac{\nu b}{2} \right) \right] c_5 \\
p_2 &=& \frac{2 a_1 a_2 \mu_1 + \mu_2^2}{a_1 a_2 \mu_1 \nu b + \mu_2^2 (1 + \nu b)} c_1 
+ \frac{1}{\nu b} \exp\left[ -\frac{1}{2} t \right] c_2 
+ \left( \frac{a_2 - 2 a_1 a_2 \mu_1}{-2 \mu_2^2 + a_2 \nu b} \right) 
\exp\left[ t \left( -\frac{1}{2} - a_1 \mu_1 - \frac{\mu_2^2}{a_2} \right) \right] c_3 \nonumber \\
&& - \exp\left[ t \left( -1 - \frac{\nu b}{2} \right) \right] c_4 \\
p_3 &=& c_1 - \exp\left[ -\frac{1}{2} t \right] c_2 
+ \exp\left[ t \left( -\frac{1}{2} - a_1 \mu_1 - \frac{\mu_2^2}{a_2} \right) \right] c_3 
+ \exp\left[ t \left( -1 - \frac{\nu b}{2} \right) \right] c_5 \\
p_3 &=& c_1 + \exp\left[ -\frac{1}{2} t \right] c_2 
+ \exp\left[ t \left( -\frac{1}{2} - a_1 \mu_1 - \frac{\mu_2^2}{a_2} \right) \right] c_3 
+ \exp\left[ t \left( -1 - \frac{\nu b}{2} \right) \right] c_4
\end{eqnarray}
\end{widetext}
where 
\begin{widetext}
\begin{eqnarray}
c_1&=&-\frac{-\mu_2^2-a_1a_2 \mu_1 \nu b-\mu_2^2 \nu b}{\left(a_2+2 a_1a_2 \mu_1+2 \mu_2^2\right) (2+\nu b)}\\
c_2&=&-\frac{\nu b}{2 (1+\nu b)}\\
c_3&=&-\frac{a_2 \left(-2 \mu_2^2+a_2 \nu b\right)}{2 \left(a_2+2 a_1a_2 \mu_1+2 \mu_2^2\right) \left(-a_2+2
a_1a_2 \mu_1+2 \mu_2^2-a_2 \nu b\right)}\\
c_4&=&-\frac{-\mu_2^2+a_1a_2 \mu_1 \nu b}{(1+\nu b) (2+\nu b) \left(a_2-2 a_1a_2 \mu_1-2 \mu_2^2+a_2 \nu b\right)}\\
c_5&=&\frac{\mu_2^2-2 a_2 \nu b+3 a_1a_2 \mu_1 \nu b+4 \mu_2^2 \nu b-3 a_2 \nu^2 b^2+2 a_1a_2 \mu_2^2 \nu b^2+2 \mu^2 \nu b^2-a_2
\nu^3 b^3}{(1+\nu b) (2+\nu b) \left(a_2-2 a_1a_2 \mu_1-2 \mu_2^2+a_2 \nu b\right)}
\end{eqnarray}
\end{widetext}

\section*{ Appendix A2} 

The Boltzmann-Gibbs entropy relation, which holds for systems far from equilibrium, is expressed as:
\begin{equation}
S[{p_{i}(t)}] = -\sum_{i=1}^3 p_{i} \ln p_{i}.
\end{equation}
 Differentiating Equation (1) with respect to time yields:
\begin{eqnarray}
\dot{S}(t) &=& \dot{e}_p(t) - \dot{h}(t) \nonumber \\
&=& \sum_{i>j}(p_{i}P_{ji} - p_{j}P_{ij}) \ln \left(\frac{p_{i}}{p_{j}}\right),
\end{eqnarray}
where $\dot{S}(t)$ represents the rate of change of entropy. The entropy extraction rate $\dot{h}_d(t)$ is given by:
\begin{eqnarray}
\dot{h}_d(t) &=& -\frac{\dot{Q}_h(t)}{T_h} + \frac{\dot{Q}_c(t)}{T_c} \nonumber \\
&=& \sum_{i>j}(p_{i}P_{ji} - p_{j}P_{ij}) \ln \left(\frac{P_{ji}}{P_{ij}}\right),
\end{eqnarray}
where $\dot{Q}_h(t)$ and $\dot{Q}_c(t)$ denote the heat absorbed from the hot reservoir and the heat transferred to the cold reservoir, respectively. The entropy production rate $\dot{e}_p(t)$ is given by:
\begin{eqnarray}
\dot{e}_p(t) &=& \sum_{i>j}(p_{i}P_{ji} - p_{j}P_{ij}) \ln \left(\frac{p_{i}P_{ji}}{p_{j}P_{ij}}\right).
\end{eqnarray}

The results of this study indicate that the rates of entropy change $\dot{S}(t)$, heat dissipation $\dot{h}_d(t)$, and entropy production $\dot{e}_p(t)$ are highest for a exponentially  decreasing temperature case. At steady state, irrespective of the temperature profile, both the entropy production and heat dissipation rates remain positive: $\dot{e}_p(t) = \dot{h}_d(t) > 0$. In the absence of a load and under isothermal conditions, both rates reduce to zero at steady state: $\dot{e}_p(t) = \dot{h}_d(t) = 0$.

 By integrating the above  rates with respect to time,  we get the corresponding extensive  thermodynamic  relations  
\begin{eqnarray} 
\Delta h_d(t) &=& \int_{t_0}^t \dot{h}_d(t) dt, \\
 \Delta e_p(t) &=& \int_{t_0}^t \dot{e}_p(t) dt, \\ 
\Delta S(t) &=& \int_{t_0}^t \dot{S}(t) dt, 
\end{eqnarray} where $\Delta S(t) = \Delta e_p(t) - \Delta h_d(t)$.   These  thermodynamic quantities  depend on network size.

The term which is related to  heat dissipation rate can be expressed as:
\begin{eqnarray}
\dot{H}_d(t) &=& \sum_{i>j} T_j (p_{i}P_{ji} - p_{j}P_{ij}) \ln \left(\frac{P_{ji}}{P_{ij}}\right) \nonumber \\
&=& \dot{E}_p(t) - \dot{S}^T(t),
\end{eqnarray}
where:
\begin{eqnarray}
\dot{E}_p(t) &=& \sum_{i>j} T_j (p_{i}P_{ji} - p_{j}P_{ij}) \ln \left(\frac{p_{i}P_{ji}}{p_{j}P_{ij}}\right), \\
\dot{S}^T(t) &=& \sum_{i>j} T_j (p_{i}P_{ji} - p_{j}P_{ij}) \ln \left(\frac{p_{i}}{p_{j}}\right).
\end{eqnarray}
Integrating the above  rates with respect to time, leads to
\begin{eqnarray}
\Delta H_d(t) &=& \int_{t_0}^t \dot{H}_d(t) dt, \\
\Delta E_p(t) &=& \int_{t_0}^t \dot{E}_p(t) dt, \\
\Delta S^T(t) &=& \int_{t_0}^t \dot{S}^T(t) dt.
\end{eqnarray}

The total internal energy $U(t)$ is given by:
\begin{equation}
U[{p_{i}(t)}] = \sum_{i=1}^3 p_{i} u_{i},
\end{equation}
with the change in internal energy:
\begin{equation}
\Delta U(t) = U[{p_{i}(t)}] - U[{p_{i}(0)}].
\end{equation}
The first law of thermodynamics is verified as:
\begin{equation}
\dot{U}[P_i(t)] = -\sum_{i>j} (p_{i}P_{ji} - p_{j}P_{ij}) (u_{i} - u_{j}) = -(\dot{H}_d(t) + fV(t)).
\end{equation}

For free energy dissipation, adapting  the isothermal case $F = U - TS$ to the non-isothermal case gives:
\begin{equation}
\dot{F}(t) = \dot{U}(t) - \dot{S}^T(t).
\end{equation}
After simplifications:
\begin{equation}
\dot{F}(t) + \dot{E}_p(t) = \dot{U}(t) + \dot{H}_d(t) = -fV(t).
\end{equation}
The change in free energy is:
\begin{equation}
\Delta F(t) = \int_{t_0}^t \left(-fV(t) - \dot{E}_p(t)\right) dt = \Delta U + \Delta H_d - \Delta E_p.
\end{equation}

\section*{Acknowledgment}
I would like to thank  Mulu  Zebene for their
constant encouragement. 

\section*{Data Availability Statement }This manuscript has no
associated data or the data will not be deposited. [Authors’
comment: Since we presented an analytical work, we did not
collect any data from simulations or experimental observations.]

\end{document}